\documentclass[aps,prl,twocolumn,superscriptaddress]{revtex4-2}
\pdfoutput=1
\usepackage{graphicx}
\usepackage{amsmath}
\usepackage{amssymb}
\usepackage{siunitx}
\usepackage[version=4]{mhchem}
\usepackage{lmodern}
\usepackage{gensymb}
\usepackage{xcolor}
\usepackage{color}
\usepackage{hyperref}

\DeclareSIUnit\gauss{G}

\newcommand{\cuone}{Cu\textsubscript{I}}
\newcommand{\cutwo}{Cu\textsubscript{II}}
\newcommand{\tnone}{\textit{T}\textsubscript{N,I}}

\newcommand{\sech}{\mathrm{sech} \,}

\begin{document}

\righthyphenmin=4
\lefthyphenmin=4

\title{High-speed antiferromagnetic domain walls driven by coherent spin waves}

\author{Kyle L. Seyler}
\affiliation{
 Department of Physics, California Institute of Technology, Pasadena, CA 91125, USA
}
\affiliation{
 Institute for Quantum Information and Matter, California Institute of Technology, Pasadena, CA 91125, USA
}
\affiliation{
 Wyant College of Optical Sciences, University of Arizona, Tucson, AZ 85718, USA
}

\author{Hantao Zhang}
\affiliation{
 Department of Electrical and Computer Engineering, University of California Riverside, Riverside, CA 92521, USA
}

\author{Daniel Van Beveren}
\affiliation{
 Department of Physics, California Institute of Technology, Pasadena, CA 91125, USA
}
\affiliation{
 Institute for Quantum Information and Matter, California Institute of Technology, Pasadena, CA 91125, USA
} 
 
\author{Costel R. Rotundu}
\affiliation{
 Stanford Institute for Materials and Energy Sciences, SLAC National Accelerator Laboratory, 2575 Sand Hill Road, Menlo Park, CA 94025, USA
}
 
\author{Young S. Lee}
\affiliation{
 Stanford Institute for Materials and Energy Sciences, SLAC National Accelerator Laboratory, 2575 Sand Hill Road, Menlo Park, CA 94025, USA
}
\affiliation{
 Department of Applied Physics, Stanford University, Stanford, CA 94305, USA
}

\author{Ran Cheng}
\affiliation{
 Department of Electrical and Computer Engineering, University of California Riverside, Riverside, CA 92521, USA
}
\affiliation{Department of Physics and Astronomy, University of California Riverside, Riverside, CA 92521, USA}
 
\author{David Hsieh}
\affiliation{
 Department of Physics, California Institute of Technology, Pasadena, CA 91125, USA
}
\affiliation{
 Institute for Quantum Information and Matter, California Institute of Technology, Pasadena, CA 91125, USA
}

\begin{abstract}
The ability to rapidly manipulate domain walls (DWs) in magnetic materials is key to developing novel high-speed spintronic memory and computing devices. Antiferromagnetic (AFM) materials present a particularly promising platform due to their robustness against stray fields and their potential for exceptional DW velocities. Among various proposed driving mechanisms, coherent spin waves could potentially propel AFM DWs to the magnon group velocity while minimizing dissipation from Joule heating. However, experimental realization has remained elusive due to the dual challenges of generating coherent AFM spin waves near isolated mobile AFM DWs and simultaneously measuring high-speed DW dynamics. Here we experimentally realize an approach where ultrafast laser pulses generate coherent spin waves that drive AFM DWs and develop a technique to directly map the spatiotemporal DW dynamics. Using the room-temperature AFM insulator \ce{Sr2Cu3O4Cl2}, we observe AFM DW motion with record-high velocities up to \SI{\sim 50}{\km\per\s}. Remarkably, the direction of DW propagation is controllable through both the pump laser helicity and the sign of the DW winding number. This bidirectional control can be theoretically explained, and numerically reproduced, by the DW dynamics induced by coherent spin waves of the in-plane magnon mode---a phenomenon unique to magnets with an easy-plane anisotropy. Our work uncovers a novel DW propulsion mechanism that is generalizable to a wide range of AFM materials, unlocking new opportunities for ultrafast coherent AFM spintronics.

\end{abstract}

\maketitle
\date{\today}

An emerging approach to drive DW motion is by using magnons\textemdash the quanta of spin waves. Thermal magnons can give rise to entropic torques in temperature gradients that move DWs towards hotter regions \cite{hinzkeDomainWall2011,kovalevThermomagnonicSpin2012,schlickeiserRoleEntropy2014,wangThermodynamicTheory2014,yanThermodynamicMagnon2015,selzerInertiaFreeThermally2016, dongesUnveilingDomain2020}, a phenomenon widely observed in different magnetic materials \cite{ashkinInteractionLaser1972,jiangDirectImaging2013,tetienneNanoscaleImaging2014,ramsayOpticalSpinTransferTorqueDriven2015,quessabHelicitydependentAlloptical2018,shokrSteeringMagnetic2019,hedrichNanoscaleMechanics2021}. In contrast, our understanding of DWs driven by \textit{nonthermal} magnons remains limited.

Theories predict that propagating magnons can drive DW motion through the transfer of spin angular momentum or linear momentum upon transmission or reflection \cite{mikhailovForcedMotion1984,hanMagneticDomainwall2009,yanAllMagnonicSpinTransfer2011,tvetenAntiferromagneticDomain2014,kimPropulsionDomain2014,wangMagnonDrivenDomainWall2015,qaiumzadehControllingChiral2018,yuPolarizationselectiveSpin2018,ohBidirectionalSpinwavedriven2019,rodriguesSpinWaveDriven2021,lanSpinWave2022, jiaoUniversalSpin2024}. In ferromagnetic systems, experimental evidence for magnonic spin-transfer torque includes the de-pinning of DWs in permalloy using spin wave bursts from colliding DWs \cite{wooMagneticDomain2017a}, light-induced magnetoelastic waves shifting DWs in iron garnets \cite{ogawaPhotodriveMagnetic2015}, and coherent magnons from microwave antennas driving DWs in metallic multilayers and iron garnet films at velocities of 10 to \SI{\sim 100}{\m\per\s} \cite{hanMutualControl2019,fanCoherentMagnoninduced2023}. Beyond ferromagnets, antiferromagnets exhibit complex spin texture dynamics \cite{gomonayAntiferromagneticSpin2018}, diverse magnon excitations \cite{hanCoherentAntiferromagnetic2023}, and high DW velocities \cite{gomonayHighAntiferromagnetic2016, shiinoAntiferromagneticDomain2016}, which may lead to powerful new approaches for fast DW manipulation \cite{tvetenAntiferromagneticDomain2014,kimPropulsionDomain2014,yuPolarizationselectiveSpin2018,qaiumzadehControllingChiral2018,rodriguesSpinWaveDriven2021, yuMagneticTexture2021}. However, nonthermal magnonic manipulation of AFM DWs remains experimentally unrealized.

Ultrafast laser pulses are a versatile way to generate nonthermal magnons in antiferromagnets. It is possible to induce intense coherent AFM spin waves with ultrafast laser light through inverse magneto-optical processes \cite{kimelUltrafastNonthermal2005,satohSpinOscillations2010,satohDirectionalControl2012,satohWritingReading2015,tzschaschelUltrafastOptical2017}, thermal effects \cite{kimelOpticalExcitation2006}, terahertz electromagnetic excitation \cite{kampfrathCoherentTerahertz2011}, and strong absorption \cite{hortensiusCoherentSpinwave2021}. However, studying the interaction between coherent spin waves and AFM DWs presents three key challenges. First, creating and detecting DWs in antiferromagnets is often more challenging than in ferrimagnets and ferromagnets due to the absence of a net moment \cite{cheongSeeingBelieving2020}. Second, no experiment has mapped the spatiotemporal evolution of light-driven DW motion, which would allow the direct measurement of fast dynamics. Ultrafast stroboscopic pump-probe experiments require the sample to return to its initial state after excitation, complicating the visualization of irreversible DW dynamics \cite{zaykoUltrafastHighharmonic2021}. Most studies of light-driven DWs either examine static before-and-after domain structure images without probing the transient behavior \cite{quessabHelicitydependentAlloptical2018} or infer spatially averaged dynamics through time-resolved diffraction experiments \cite{jangidExtremeDomain2023}. Third, ultrafast optical excitation can affect AFM DWs through more trivial thermal effects \cite{quessabHelicitydependentAlloptical2018,shokrSteeringMagnetic2019,hedrichNanoscaleMechanics2021} or non-thermal photoinduced magnetic anisotropy \cite{stupakiewiczLightinducedMagnetic2001a,kirilyukUltrafastOptical2010}, which must be distinguished from coherent magnonic effects.

\begin{figure}[!htb]
\includegraphics{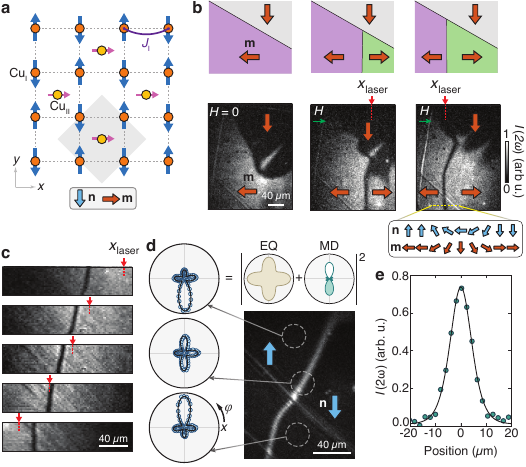}
\caption{\label{fig:m1} \textbf{Creation and positioning of wide antiphase Néel DWs.} \textbf{a}, Crystal and magnetic structure of the \ce{Cu3O4} plane of \ce{Sr2Cu3O4Cl2} below \tnone. \textbf{b}, Creation of an antiphase DW at a \ang{90} DW. The bottom row shows SHG images of the process at different $H$ and horizontal laser positions ($x_\text{laser}$), while the top row shows a schematic of the domain configuration. The resulting antiphase DW configuration is depicted in the bottom right. The imaging laser is horizontally polarized (along $x$ axis). \textbf{c}, SHG images of an antiphase DW at different $x_\text{laser}$. \textbf{d}, SHG image of antiphase DW with vertically polarized laser. Rotational anisotropy polar plots of the SHG intensity (for \textit{P}-polarized input and output electric field polarization measured as a function of the scattering plane angle $\varphi$) are shown for selected locations. $\varphi=0$ corresponds to the $x$ direction. The solid lines are fits to a coherent superposition of crystallographic electric quadrupole (EQ) and AFM-induced magnetic dipole (MD) SHG processes, as described in ref.~\cite{seylerDirectVisualization2022}. The EQ and MD processes shown in the figure represent the \textit{P}\textsubscript{in}-light-induced nonlinear polarization projected along \textit{P}\textsubscript{out}. Filled and unfilled lobes indicate opposite phases. The DW SHG pattern is fit to a three-domain averaged expression, as described in Supplementary Section 1. \textbf{e}, Line cut of the SHG image intensity perpendicular to an antiphase DW. The solid line shows a fit to the SHG intensity of a DW profile (see Methods) with an extracted DW width $\pi\mathit{\lambda}=\SI{8.98\pm 0.22}{\um}$}.
\end{figure}

Here we overcome these challenges and directly capture the ultrafast-light-induced motion of an AFM DW. Our material of choice, the square-lattice Mott insulator \ce{Sr2Cu3O4Cl2}, hosts in-plane Cu spins with strong AFM exchange interactions similar to high-temperature cuprate superconductor parent compounds, where the \cuone{} sublattice has $J_{\textrm I} \approx \SI{130}{\meV}$ with a Néel temperature of $\tnone \approx \SI{380}{\kelvin}$ (Fig.~\ref{fig:m1}a). However, \ce{Sr2Cu3O4Cl2} is unique in that it contains an additional Cu ion (\cutwo) at the center of every other square plaquette, breaking the equivalence of neighboring \cuone{} sites and inducing a weak in-plane ferromagnetic moment $\mathbf{m}$ locked perpendicular to the Néel vector $\mathbf{n}$ \cite{chouFerromagneticMoment1997}. Previous work has shown that optical second-harmonic generation (SHG) rotational anisotropy, which is sensitive to magnetic point group symmetries, can detect $\mathbf{m}$ and $\mathbf{n}$, enabling direct visualization of AFM domains and DWs \cite{seylerDirectVisualization2022}. The locations of the \ang{90} DWs were largely fixed by a built-in spatially dependent uniaxial anisotropy. In ambient field conditions, \ang{180} (antiphase) DWs were not observed, due to poling by the Earth’s magnetic field during cooling.

\section{Creating and characterizing antiphase Néel-type AFM DW\MakeLowercase{s}}

We first demonstrate controlled creation and positioning of antiphase AFM DWs using applied magnetic fields and laser heating. Polarized wide-field SHG images exhibit distinct bright and dark regions that correspond to magnetic domains with perpendicular $\mathbf{m}$ orientations (Fig.~\ref{fig:m1}b), as described in ref.~\cite{seylerDirectVisualization2022}. Initially, at zero applied magnetic field ($H=0$), only two domain states are present ($\mathbf{m}$ along $-x$ and $-y$). By applying $H$ anti-parallel to the $-x$ domain, a third domain state (with $\mathbf{m}$ along $+x$) becomes more favorable. Eventually, at high-enough $H$, an antiphase DW would form and propagate through the sample to flip the $-x$ domain to $+x$. To stabilize this antiphase DW within the sample, we fixed $H$ just below this threshold value for antiphase DW formation ($\mu_0 H \approx \SI{6}{\gauss}$) and scanned the imaging laser across the \ang{90} wall. Local laser heating reduces the coercive field, allowing the formation of an antiphase DW that becomes spatially trapped near the imaging laser center. The antiphase DW manifests as a vertical dark line in the image. Moreover, it can be dragged to different positions by translating the imaging laser (Fig.~\ref{fig:m1}c). The DW prefers the laser center due to an entropic torque caused by the laser-induced Gaussian temperature profile.

Next, we characterize the width, type, and winding number of the antiphase DWs. In Fig.~\ref{fig:m1}b and \ref{fig:m1}c, the dark DW line could arise from destructive interference of SHG at the boundary of the time-reversed domain states \cite{fiebigProbingFerroelectric2002}. To rule this out, we rotated the imaging laser polarization by \ang{90}, causing the previously bright domain states to appear dark. Surprisingly, the antiphase DW shows comparably bright SHG (Fig.~\ref{fig:m1}d) relative to the single domains in Fig.~\ref{fig:m1}b and \ref{fig:m1}c, implying that the DW width must be large compared to the SHG wavelength (\SI{400}{\nm}) and that SHG interference between neighboring domains is negligible. A line cut across the DW reveals a DW width $\pi\mathit{\lambda}\approx\SI{9}{\um}$, where $\mathit{\lambda}$ is the width parameter, which is larger than both our diffraction limit (\SI{\sim 1}{\um}) and the DW widths in typical antiferromagnets such as \ce{NiO} and \ce{Cr2O3} (\SIrange{10}{100}{\nm}) \cite{schmittIdentifyingDomainwall2023,wornleCoexistenceBloch2021}. The wide antiphase DWs are a consequence of the large $J_{\textrm I}$ combined with relatively weak in-plane easy-axis anisotropy ($K$) \cite{chouFerromagneticMoment1997, parksMagnetizationMeasurements2001} since the width scales as $\sqrt{J_{\textrm I}/K}$.

\begin{figure*}[!htb]
\includegraphics{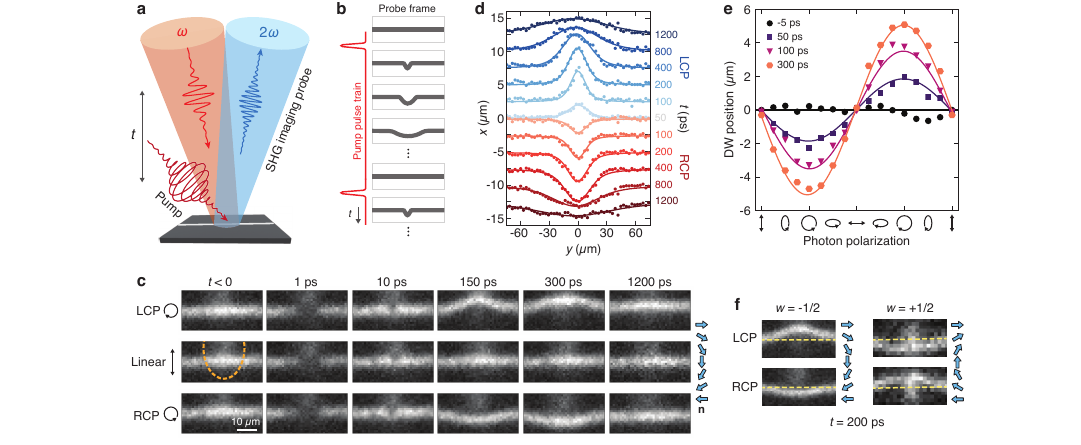}
\caption{\label{fig:m2} \textbf{Observation of helicity-dependent light-driven DW motion.} \textbf{a}, Schematic of pump-probe SHG imaging experiment. The pump beam is focused obliquely on the sample at \ang{\sim 10} angle of incidence, while the SHG imaging probe is near-normal incidence. \textbf{b}, Schematic of the DW position at different times before and after each pulse of pump excitation train. \textbf{c}, SHG images of an antiphase DW at selected pump-probe time delays for linear, left circular, and right circular pump polarizations. The dashed oval indicates the pump excitation spot, which is offset from the DW center. The DW configuration is displayed on the right. \textbf{d}, Extracted dynamics of the DW shape for left and right circular pump. The solid curves are fits to a Voigt profile. The data at different delays are vertically offset for clarity. \textbf{e}, Pump photon polarization dependence of the maximum DW position (at $x=0$) for different time delays. The horizontal axis runs from $\theta=\ang{-90}$ to $\theta=\ang{90}$, where $\theta$ is the linear polarization angle before entering a quarter-wave plate with fixed fast axis at \ang{0}. The solid curves are fits proportional to $\sin(2\theta)$. \textbf{f}, SHG images at $t=\SI{200}{\ps}$ for different pump photon helicities and DW winding numbers $w$. The dashed lines indicate the initial DW position at $t < 0$. The additional intensity outside the DW location arises due to scattered SHG from the pump beam as well as a contribution from spin wave precession for $t>0$ (see main text). The vertical scale of each image is \SI{20}{\um}.}
\end{figure*}

The large DW width enables us to determine the orientation of $\mathbf{n}$ within the DW using scanning SHG rotational anisotropy. In single-domain regions, SHG patterns exhibit a strong lobe oriented along $-\mathbf{n}$, consistent with prior results \cite{seylerDirectVisualization2022} (Fig.~\ref{fig:m1}d). In contrast, the DW region shows a pattern characteristic of a superposition of the two single-domain states (Supplementary Section 1). Notably, the $\varphi = \ang{0}$ lobe is more intense than the $\varphi = \ang{180}$ lobe, indicating that $\mathbf{n}$ is oriented along $-x$ at the DW center. These results confirm that the DW is Néel-type (Fig.~\ref{fig:m1}b inset), where the spins rotate within the sample plane, perpendicular to the DW normal, as expected from the easy-plane anisotropy of the parent cuprates. We may characterize its anti-clockwise sense of spin rotation through a winding number $w = -1/2$, which gives the number of times the spins wrap around a circle in the clockwise direction ($w = \frac{1}{2\pi} \int_{-\infty}^{+\infty} \partial_x \phi(x) \, dx$, where $\phi(x)$ is the in-plane Néel vector angle at position $x$ along the wall). Our SHG measurements can therefore fully characterize the width, type, and winding number of the AFM DWs.

\section{Spatiotemporal mapping of helicity-dependent DW dynamics}

We now proceed to study the dynamics after ultrafast optical excitation on the AFM DWs using time-resolved pump-probe SHG imaging. Figure~\ref{fig:m2}a illustrates the experimental configuration. We used \SI{100}{\fs} pump pulses (fluence \SI{8}{\milli\joule\per\cm^2}) at \SI{1.55}{\eV}, below the \SI{1.86}{\eV} charge gap \cite{shanDynamicMagnetic2024}, focused at slightly oblique incidence to an elliptical spot on the DW, while time-delayed wide-field SHG imaging pulses served as the probe (see Methods). The laser repetition rate is \SI{100}{kHz} with SHG imaging integration times of ${\sim} 10^5$ seconds, so each final image consists of an average of ${\sim}10^{10}$ individual pump excitation events. Therefore, in order to resolve the spatiotemporal dynamics, the DW must return to its original state before the next pulse in the train arrives (Fig.~\ref{fig:m2}b). Here we realize reversible DW motion through the combined effects of DW surface tension and the photothermal trapping potential of the large SHG imaging beam (Fig.~\ref{fig:m1}c), which generate restoring forces against localized perturbations. Under either linearly or circularly polarized pumping, the DW SHG intensity strongly decreases soon after time zero ($t = \SI{1}{\ps}$). This indicates ultrafast suppression of the 3D long-range AFM order due to photothermal heating, though in-plane short-range order can still persist \cite{shanDynamicMagnetic2024}. Within \SI{10}{\ps}, the DW almost fully recovers, consistent with previously measured spin dynamics in \ce{Sr2Cu3O4Cl2} \cite{shanDynamicMagnetic2024}. At later times, we observe no subsequent changes in the SHG images for linearly polarized pumping. In contrast, left or right circularly polarized pump pulses induce striking changes, driving the DW locally up or down, respectively, followed by a slower relaxation back to the equilibrium DW position. The crisp DW images demonstrate that the motion is reversible. We extract the DW shape by fitting to vertical line cuts across the wall and plotting the center positions as shown in Fig.~\ref{fig:m2}d. This illustrates how the DW not only propagates up or down normal to itself but also moves outwards to the left and right, tens of microns beyond the pump laser pulse spot. Interestingly, these complex DW dynamics occur well after the \SI{100}{\fs} pump light has interacted with the sample, as visualized in \href{https://youtube.com/shorts/5QBzSEGMKoA?feature=share}{Supplementary Videos 1} and \href{https://youtu.be/wLNATZVRGGk}{2}.

\begin{figure}[!tb]
\includegraphics{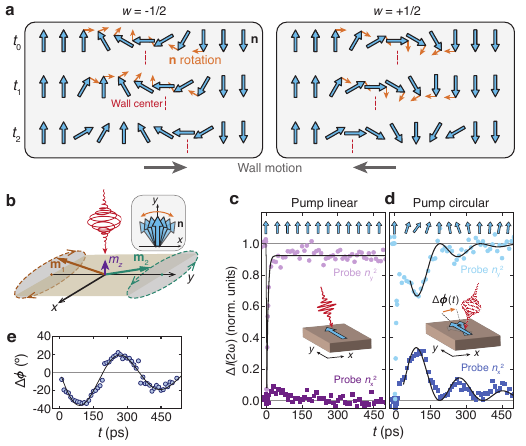}
\caption{\label{fig:m3} \textbf{Mechanism of light-driven DWs.} \textbf{a}, Schematic of Néel-type DW at different times ($t_0$, $t_1$, $t_2$) after clockwise rotations of $\mathbf{n}$ for negative (left panel) and positive (right panel) winding number. \textbf{b}, Schematic of coherent in-plane magnon induced by a circularly polarized laser pulse. $\mathbf{m}_1$ and $\mathbf{m}_2$ indicate the sublattice moments and $m_z$ is the out-of-plane moment induced by laser. The Néel vector $\mathbf{n} = (\mathbf{m}_2 – \mathbf{m}_1)/2$ initially rotates in the clockwise direction and its full trajectory is a back-and-forth oscillation in the $xy$ plane (inset). \textbf{c}, SHG transients for linear pump excitation with the probe beam linearly polarized in the $\hat{x}$ (circular markers) or $\hat{y}$ (square markers) direction, giving SHG proportional to $n_y^2$ or $n_x^2$ respectively. The black curves are guides to the eye. \textbf{d}, Same as \textbf{c} but with circularly polarized pump. The black curves are fitted to the square of a damped sine wave: $I_\text{SHG}=I_0(\sin{(2\pi f t+\phi_0)}e^{-t/\tau})^2$, where $I_0$ is the intensity, $f$ is the frequency of oscillation, $\phi_0$ is a phase offset, and $\tau$ is the decay time constant. \textbf{e}, Extracted change in Néel vector angle, $\Delta\phi(t)$, using the case of circular pump with $\hat{x}$ probe in \textbf{d}. Since the SHG cannot distinguish between positive and negative $\Delta\phi$, we have chosen the signs of the data to give agreement with $f\approx \SI{2.8}{\GHz}$ found in \textbf{d}. The black curve is a fit to a damped sine wave.}
\end{figure}

\section{Coherent spin waves drive the DW motion}

To uncover the origin of the light-induced DW motion, we explored different pump laser parameters and DW configurations. First, we systematically adjusted the pump polarization between linear, elliptical, and circular states, finding maximal DW displacement for circularly polarized light, with opposite helicities driving motion in opposite directions (Fig.~\ref{fig:m2}d). Photoinduced magnetic anisotropy can thus be ruled out, as it should occur under linear but not circular polarization \cite{hansteenNonthermalUltrafast2006}. In addition, the absence of DW displacement in the linear pump case implies that transient photothermal gradients do not drive the motion \cite{quessabHelicitydependentAlloptical2018}. Further evidence against a heating effect comes from the observation that, for a fixed helicity, the direction of DW motion is independent of whether the pump laser spot is offset to one side or the other of the DW (Supplementary Section 2). One might also propose that an effective in-plane photo-induced magnetic field \cite{kimelUltrafastNonthermal2005} generated from an oblique incidence circularly polarized pump laser could move the DW by coupling to its weak in-plane ferromagnetism, but we found no dependence on the pump scattering plane angle (Supplementary Section 3). Next, we examined how the DW configuration influences its motion. By starting with different initial \ang{90} DW states (Fig.~\ref{fig:m1}b), we could prepare other antiphase DW configurations that possess time-reversed DW spins or opposite sign of $w$ (i.e., whether the spins
rotate clockwise or anti-clockwise). We observed that changing the sign of $w$ reversed the DW direction for a fixed pump helicity, as shown in Fig.~\ref{fig:m2}f. On the other hand, when the winding number was fixed, time-reversing the DW state (i.e., flipping each spin) did not affect its motion (Supplementary Section 4). These results establish that the direction of DW motion is determined by a combination of the DW winding number sign and pump helicity.

This unique dependence on pump helicity and wall winding suggests an intuitive picture for the initiation of DW motion. Consider a Néel wall in an easy-plane antiferromagnet, where $\mathbf{n}$ can either rotate clockwise or anti-clockwise within the two-dimensional plane as one moves from left to right across the wall (Fig.~\ref{fig:m3}a). If there is a local clockwise rotation of $\mathbf{n}$ within the DW, the wall center position will move to the left (right) for positive (negative) winding number, while anti-clockwise rotation of $\mathbf{n}$ produces the opposite motion. Ultrafast laser pulses can induce such a rotation of $\mathbf{n}$ through the excitation of coherent spin waves (the in-plane mode) \cite{satohSpinOscillations2010}. Circularly polarized light normally incident upon the sample injects spin angular momentum along $\pm\hat{z}$, with the sign determined by the helicity. This angular momentum cants the \cuone{} sublattice moments ($\mathbf{m}_1$ and $\mathbf{m}_2$) toward the $\pm z$ direction, causing an instantaneous non-collinearity between $\mathbf{m}_1$ and $\mathbf{m}_2$, which generates an in-plane ($xy$) rotation of $\mathbf{n}$ through the exchange interaction (Fig.~\ref{fig:m3}b) \cite{chengUltrafastSwitching2015}. Consequently, the pump helicity sets the initial rotation direction of $\mathbf{n}$, thereby setting the initial DW motion. An effective view is that the photon spin angular momentum is transferred to the DW and drives its motion, forming a spin current. As $\mathbf{n}$ reverses directions, a slowdown and a reversal of the wall direction occurs. Notably, we observe no oscillations in the DW position. This is likely due to the strong damping of $\mathbf{n}$ oscillations, particularly near the DW where energy is transferred to the wall motion.

To search for evidence of light-induced coherent AFM spin waves, we performed time-resolved SHG intensity measurements on a single magnetic domain region. The desired magnon mode corresponds to $\mathbf{n}$ oscillations in the $xy$ plane, characterized by an exchange of intensity between its $x$ and $y$ components ($n_x$ and $n_y$). Normal incidence SHG intensity for linearly polarized excitation along $\hat{x}$ ($\hat{y}$) is proportional to $n_y^2$ ($n_x^2$), allowing us to measure light-induced rotational changes of $\mathbf{n}$. Figure~\ref{fig:m3}c shows the normalized change in SHG intensity after linearly polarized pumping on a domain with $\mathbf{n} = n_y \hat{y}$. As expected, $n_y$ is suppressed on ultrafast timescales and recovers within \SI{\sim 10}{\ps}, while $n_x$ shows no appreciable changes. In contrast, for the same scenario under circularly polarized excitation (Fig.~\ref{fig:m3}d), $\mathbf{n}$ oscillates in the plane at \SI{\sim 2.8}{\GHz}, corresponding to the low-frequency in-plane magnon mode depicted in Fig.~\ref{fig:m3}b that has previously been measured by electron spin resonance \cite{katsumataDirectObservation2001}. Using the SHG data in Fig.~\ref{fig:m3}d, and assuming $\dot{\phi}(t=0)<0$, we can reconstruct the trajectory of $\mathbf{n}$ and find a maximum initial rotation angle of \ang{\sim 33} (Fig.~\ref{fig:m3}e). Since the excited spin waves are quasi-uniform (optical pump spot size \SI{\sim 30}{\um}), no appreciable spin wave propagation outside the pumped region is observed. Even though the Néel vector is initially suppressed by the pump, coherent magnons can still be launched due to the presence of in-plane magnetic correlations.

\section{Quantitative analysis of high-speed DW motion and comparison to simulations}
\begin{figure}[t]
\includegraphics{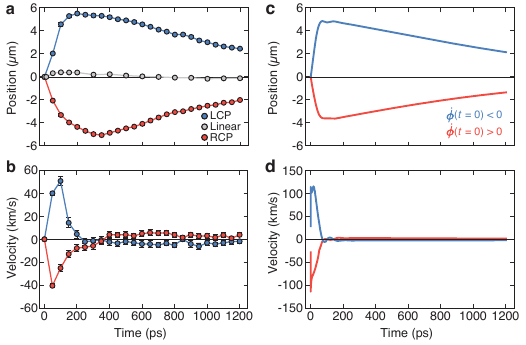}
\caption{\label{fig:m4} \textbf{Quantitative extraction of DW dynamics and comparison to simulation.} \textbf{a}, \textbf{b}, Position (\textbf{a}) and velocity (\textbf{b}) of the DW center (at the location of maximum displacement) over time for different pump polarizations. Error bars represent one standard error for the fitted DW positions and the corresponding propagated error for calculated velocities. \textbf{c}, \textbf{d}, Simulated time dependence of the DW position (\textbf{c}) and velocity (\textbf{d}) for positive and negative initial angular velocities of the Néel vector.}
\end{figure}

To quantify the light-driven DW motion, we analyzed its position and velocity over time at the location of maximum DW displacement ($x=0$ in Fig.~\ref{fig:m2}d). Figure~\ref{fig:m4}a (\ref{fig:m4}b) shows the temporal evolution of the wall center position (velocity) for different pump helicities. The DW position rapidly shifts in the first \SI{\sim 100}{\ps} with a maximum velocity of \SI{\sim 50}{\km\per\s} (Fig.~\ref{fig:m4}b), slows down from \SIrange{100}{200}{\ps} reaching a maximum displacement of \SI{\sim 6}{\um}, and then reverses direction back to equilibrium at \SI{\sim 5}{\km\per\s}. The \SI{\sim 100}{\ps} timescale of the fast outward wall motion is similar to the timescale of the initial $\mathbf{n}$ rotation identified in Fig.~\ref{fig:m3}e. Furthermore, reducing the pump fluence does not appreciably change this timescale (Supplementary Section 5), although it decreases the maximum velocity and position. These observations confirm the causal chain: the ultrafast optical pulse rapidly excites coherent spin waves, which subsequently propel the DW over a significantly longer timescale. The slight asymmetry between left and right circularly polarized cases is likely due to the small spatial offset between pump laser spot and DW (Fig.~\ref{fig:m2}c). The measured DW velocity reaches an order of magnitude greater than the speed of longitudinal sound waves in cuprates \cite{miglioriElasticConstants1990,abd-shukorUltrasonicElastic2018}, but about four times smaller than the large magnon group velocity of \ce{Sr2Cu3O4Cl2} (see Methods), which sets the upper limit for DW velocity. Notably, the velocity exceeds reports of coherent-magnon-driven DW velocities ($\SI{\leq 1}{\km\per\s}$) in ferromagnetic films \cite{hanMutualControl2019,fanCoherentMagnoninduced2023}.

To understand the microscopic mechanisms behind the DW motion, we performed one-dimensional numerical simulations based on the established mean-field magnetic energy of \ce{Sr2Cu3O4Cl2} \cite{chouFerromagneticMoment1997, kimNeutronScattering2001, gomonayTheoryMagnetization2011}, which captures the most essential features of the observed DW motion. In this simplified picture, the DW profile is described by a single variable, the in-plane angle $\phi=\phi(x,t)$, which satisfies (see Methods):
\setlength{\abovedisplayskip}{4pt}
\begin{align}
    \partial^{2}_{t} \phi - v_{m}^{2} \partial^{2}_{x}\phi - & \frac{v_{m}^{2}}{2 \lambda^{2}} \sin 2 \phi + \frac{\alpha v_m}{a} \partial_{t} \phi \nonumber \\
    &+ \frac{U}{2\lambda^2} \sech^2 \frac{x}{\lambda}(\phi - \phi_c)= 0
\end{align}
where $v_m$ is the magnon velocity, $\alpha$ is the Gilbert damping, and $a$ is the lattice constant. The last term accounts for the DW trapping potential from DW tension and photothermal heat gradients, where $U$ is a phenomenological constant and $\phi_c$ is the value of $\phi$ at the DW center. The spin wave excitation is modeled by setting the initial in-plane spin velocities $\dot{\phi}(x,t=0)$ over a Gaussian spatial profile with a sign dependent on the helicity. Solving the effective dynamical equation for realistic material parameters and initial conditions, we were able to reproduce the key experimental features including the helicity dependence, winding number dependence, micron-scale net displacement, fast motion, and slow recovery to equilibrium as shown in Fig.~\ref{fig:m4}c and \ref{fig:m4}d (see also Supplementary Section 6, \href{https://youtu.be/3JurysRfQSI}{Supplementary Video 3}). Our theory and simulations show that the spatial profile of the pump beam is not critical to the observed DW motion. Even under spatially uniform driving, the intrinsic inhomogeneity of the DW itself generates a non-uniform angular acceleration of the spins that propels the DW (Supplementary Section 7). Because our simulation is limited to one dimension, we are unable to capture the transverse motion along the DW line. The observed dynamics are further complicated by the fact that the DW is three-dimensional membrane that extends into the sample bulk, and optical driving is strongest near the surface. In contrast to prior theoretical works on magnon-driven AFM DWs that have focused on easy-axis antiferromagnets \cite{tvetenAntiferromagneticDomain2014,kimPropulsionDomain2014, selzerInertiaFreeThermally2016,qaiumzadehControllingChiral2018,yuPolarizationselectiveSpin2018}, our findings indicate that spin waves can also drive fast DW motion in easy-plane antiferromagnets. Our proposed driving mechanism is unique in that it requires the spin waves to be coherent, with the fast motion in \ce{Sr2Cu3O4Cl2} assisted by the large-angle character of the spin wave excitations.

In conclusion, we established light-induced coherent in-plane AFM spin waves as an effective mechanism for driving fast directional DW motion, which should be applicable to other magnets with easy-plane anisotropy. Optimizing the driving protocol with different pump energies may enable even higher AFM DW velocities and the study of relativistic effects \cite{haldaneNonlinearField1983a, baryakhtarDynamicSolitons1984, kimPropulsionDomain2014, gomonayHighAntiferromagnetic2016,shiinoAntiferromagneticDomain2016,yuanClassificationMagnetic2018a, carettaDomainWalls2024}. Furthermore, our results provide an unprecedentedly clear and rich view of DW dynamics that can be used to further refine theoretical calculations beyond simple one-dimensional models \cite{chengMagneticDomain2019}. We anticipate that our findings will stimulate further exploration of how different types of coherent AFM spin waves interact with spin textures in antiferromagnets more broadly.

\section{Methods}

\subsection{Sample growth}
\ce{Sr2Cu3O4Cl2} crystals were grown by an optimized method of slow cooling from the melt \cite{noroMagneticProperties1994}. Quantities of \ce{SrO}, \ce{SrCl2}, and \ce{CuO} powders were mixed in a 1:1:3 stoichiometric ratio and placed in a large high form alumina crucible. The mix was gradually heated in air \SI{1030}{\degreeCelsius}, dwelled for \SI{5}{\hour}, then cooled to \SI{900}{\degreeCelsius} at a rate of \SI{2}{\degreeCelsius\per\hour}. Placing the crucible in a slight temperature gradient (off-center of the hot chamber of the box furnace) resulted in a cm-sized plate-like single crystal. The samples showed excellent stability in air. X-ray diffraction, Laue X-ray, and low-temperature magnetization measurements were used to verify high sample quality. The samples were affixed to an oxygen-free high-thermal-conductivity copper mount using a small amount of silver epoxy and then cleaved before measurement to leave clean surfaces parallel to the \ce{Cu3O4} (001) planes.

\subsection{SHG measurements}
\textit{SHG rotational anisotropy}: SHG rotational anisotropy measurements were carried out using a fast-rotating scattering plane based technique \cite{harterHighspeedMeasurementRotational2015} with laser pulses delivered by a Ti:sapphire amplifier (800 nm fundamental wavelength, \SI{100}{\fs} pulse duration, \SI{100}{\kHz} repetition rate). The beam diameter was \SI{40}{\um} with a fluence of \SI{3}{\milli\joule\per\cm^2}.

\textit{SHG imaging}: Wide-field SHG imaging was performed using the same laser source as that used for rotational anisotropy measurements. The imaging was performed under normal incidence, with linearly polarized excitation, a fluence of \SI{2.3}{\milli\joule\per\cm^2}, and a full-width-at-half-maximum beam diameter of \SI{210}{\um}, corresponding to \SI{80}{\mW} of average power. The reflected SHG was collected by an achromatic doublet objective lens and imaged onto a cooled charge-coupled device camera. Using the form of DW profile given in Eq.~\eqref{eq:static}, the resulting SHG intensity profile is $I(x)=I_0 [\sech{((x-x_0)/\lambda)}]^2$, where $I_0$ is the maximum SHG at the DW center and $x_0$ is the DW center. This equation was used to extract the DW parameters from the experiments. 

\textit{Time-resolved SHG imaging}: Experiments were performed by splitting off 800 nm light from the same laser source to produce the pump beam. The pump beam was sent through a delay line before reaching the objective doublet lens spatially offset from the principal axis, after which it was focused onto the sample at a \ang{10} angle of incidence. The pump beam spot on the sample was elliptically shaped with a full-width-at-half-maximum of \SI{15}{\um} and \SI{10}{\um} for the major and minor axes of the ellipse respectively. The reflected pump beam was physically blocked to reduce the amount of scattered pump SHG light that reached the camera. A small amount of SHG light generated by the pump is detected by the camera, which allows us to locate the pump beam relative to the DW. The pump and probe fluences were \SI{8}{\milli\joule\per\cm^2} and \SI{2.3}{\milli\joule\per\cm^2} respectively. The time-resolved two-dimensional DW dynamics (Fig.~\ref{fig:m2}d) were quantitatively extracted by fitting to vertical line cuts of the DW images in Fig.~\ref{fig:m2}c. The fit uncertainties are given as 1 standard error. The location of maximum DW motion ($x=0$ in Fig.~\ref{fig:m2}d) was used for the one-dimensional dynamics reported in Fig.~\ref{fig:m2}e and Fig.~\ref{fig:m4}). The time-resolved SHG intensity traces in Fig.~\ref{fig:m3}c and d were acquired from time-resolved SHG images on a single-domain region. The SHG intensities were calculated from the average intensity in an approximately \SI{6}{\um} by \SI{6}{\um} box at the center of the pump excitation profile.

\subsection{AFM DW dynamics theory and simulation}
To characterize the most essential DW dynamics observed in \ce{Sr2Cu3O4Cl2}, we adopt a simplified effective 1D model by ignoring the transverse motion along the DW line, using a normalized Néel vector $\mathbf{n}(x,t)$ to describe the longitudinal DW motion (along $x$). Based on prior investigations~\cite{gomonayTheoryMagnetization2011, chouFerromagneticMoment1997, kastnerFielddependentAntiferromagnetism1999, kimNeutronScattering2001, tvetenIntrinsicMagnetization2016, shiinoAntiferromagneticDomain2016}, we construct a phenomenological free energy in the continuum as
\begin{align}
\label{eq:magnetic_energy}
    E = &S^2\int dx \left[ \eta \mathbf{m}^{2} + \beta \left(\partial_{x} \mathbf{n} \right)^2 + J_{\textrm I} \mathbf{m} \cdot \partial_x\mathbf{n} \right.  \nonumber \\
    &\left. + \left( J_{\rm pd} \mathbf{M}_{F}\sigma_{1}\mathbf{n} - J_{\rm av} \mathbf{m} \cdot \mathbf{M}_{F} + K_{\perp}n_{z}^{2}\right)/a\right],
\end{align}
where $S$ is the spin quantum number of \cuone{} atoms, $\mathbf{m}$ and $\mathbf{M}_{F}$ are the dimensionless vectors of the local magnetization arising from the \cuone{} and \cutwo{} sublattices, respectively. $\eta = J_{\textrm I}/a$ and $\beta = J_{\textrm I}a/2$ are the homogeneous and inhomogeneous stiffness constants where $a$ is the lattice constant and $J_{\textrm I}>0$ is the nearest-neighbor Heisenberg exchange coupling (illustrated in Fig.~\ref{fig:m1}a). Besides a dominant hard-axis anisotropy $K_\perp>0$ suppressing the out-of-plane canting of $\mathbf{n}$, we also include $J_{\rm av}>0$ and $J_{\rm pd}>0$ as the isotropic and anisotropic pseudo-dipolar interactions affecting the in-plane components, where $\sigma_1$ is the Pauli matrix acting on the $x$--$y$ coordinates. Here we use $\sigma_1$ instead of $\sigma_3$ (which appears in previous studies) because our convention sets the domains as $\phi\sim\pm\pi/2$ rather than $\pi/4$ and $3\pi/4$. In real materials, there could exist additional anisotropy mechanisms renormalizing $J_{\rm pd}$, but that will not change the formalism or the phenomenology.

When $K_{\perp}$ dominates other anisotropies, the dynamics of $\mathbf{m}$, $\mathbf{n}$ and $\mathbf{M}_{F}$ mainly involves their in-plane components, thus to a good approximation we can parameterize $\mathbf{n}$ by a single variable $\phi$ such that $\mathbf{n} \approx \cos \phi \hat{\mathbf{x}} + \sin \phi \hat{\mathbf{y}}$. The in-plane dynamics of $\mathbf{n}$ is inherently related to the low-frequency (acoustic) mode of magnons~\cite{chengUltrafastSwitching2015}, for which $\mathbf{M}_{F}$ is able to adiabatically follow the instantaneous motion of $\mathbf{n}$. In this regard, $\mathbf{M}_{F}$ will be essentially locked orthogonally to $\mathbf{n}$. Since $\mathbf{m}$ is also almost in-plane and $\mathbf{m}\perp\mathbf{n}$ by definition, $\mathbf{M}_{F}$ is collinear with $\mathbf{m}$ at all locations, ergo $J_{\rm av}\mathbf{m}\cdot\mathbf{M}_{F}$ reduces to a constant and can be ignored. The same approximation renders the term $ J_{\rm pd} \mathbf{M}_{F}\sigma_{1}\mathbf{n}$ proportional to $\cos2\phi$, serving as an effective in-plane easy-axis anisotropy. Correspondingly, the free energy $E$ becomes a functional of $\phi(x)$.

Such a quasi-1D AFM texture can be described by the action
$\mathcal{S}[\phi]=\mathcal{S}_W[\phi,\partial_t\phi] + \int dt E[\phi,\partial_x\phi]$ where $\mathcal{S}_W$ is the Wess-Zumino-Witten term~\cite{fradkin2013field}. By integrating out $\mathbf{m}$ using the path integral approach, we obtain the effective action as a functional of $\phi$ as
\begin{align} 
\label{eq:effective_action}
 &\mathcal{S}_{\rm eff}[\phi] = \int dtdx \mathcal{L}_{\rm eff} = S^2\int  dt dx \notag\\
 &\quad \left[\frac{\hbar^2}{J_{\textrm I}aS^2}(\partial_t\phi)^2 - \frac{J_{\textrm I}a}4 (\partial_{x}\phi)^{2} - \frac{J_{\rm pd}M_F}{a} \cos 2\phi \right],
\end{align}
where $\hbar$ is the reduced Planck constant. The Berry phase terms of $\mathbf{n}$ and $\mathbf{M}_F$ have been discarded as we are focusing on the local and in-plane dynamics of $\mathbf{n}(x,t)$. By applying the Euler-Lagrange equation of $\mathcal{L}_{\rm eff}$ in the presence of Rayleigh's dissipation (density) function $\mathcal{R}=\alpha S/(2a)(\partial_t\phi)^2$~\cite{kamraGilbertDamping2018} to account for the Gilbert damping $\alpha$, we obtain
\begin{align} \label{eq:dynamical_eq}
    \partial^2_t\phi - v_m^2\partial^2_x\phi - \frac{v_m^2}{2\lambda^2} \sin2\phi + \frac{\alpha v_m}{a} \partial_t\phi = 0,
\end{align}
where $v_m=J_{\textrm I}aS/(2\hbar)$ is the magnon velocity and $\lambda=a\sqrt{J_{\textrm I}/(8J_{\rm pd}M_F)}$ is the DW width parameter. Under the asymptotic boundary conditions $\phi(\pm\infty)$, the static solution to Eq.~\eqref{eq:dynamical_eq} is obtained as
\begin{align}
 \phi_0(x)=2\arctan e^{\pm x/\lambda} \pm \frac{\pi}2, \label{eq:static}
\end{align}
where the $+$ ($-$) sign corresponds to the $w=-1/2$ ($+1/2$) DW illustrated in Fig.~\ref{fig:m3}a.

Equation~\eqref{eq:dynamical_eq} is insufficient to capture the backward DW motion following its initial outward drive, which calls for a trapping potential to provide an effective restoring force. The physical origin of such a trapping potential could be the surface tension of the DW line and the photo-thermal heating from the laser. To phenomenologically construct a trapping potential, we consider a small virtual displacement of the DW center $\Delta x$ so that the static profile $\phi_0(x)$ becomes $\phi_0(x-\Delta x)$. Correspondingly, the local change of $\phi$ is proportional to $\partial_x\phi_0$ [\textit{i.e.}, $\phi_0'(x)$], thus a harmonic trapping potential should be proportional to $\phi_0'^2$. However, $\phi_0'^2$ alone cannot affect the field dynamics because it contributes a constant to the free energy. The potential should be locally proportional to $(\phi - \phi_{c})^{2}$ with $\phi_{c}$ the value of $\phi$ at the DW center, reminiscent of the Klein-Gordon theory. In Fig.~\ref{fig:m3}a, we have $\phi_c=0$ ($\pi$) for the $w=+1/2$ ($-1/2$) DW. Therefore, the trapping potential assumes the form $U\phi_{0}'^{2} (\phi - \phi_{c})^{2}$ where $U$ is a phenomenological constant.

After incorporating the trapping potential into the free energy, we can rederive Eq.~\eqref{eq:dynamical_eq} as
\begin{align} \label{eq:dynamical_eq_re}
    \partial^{2}_{t} \phi - v_{m}^{2} \partial^{2}_{x}\phi - & \frac{v_{m}^{2}}{2 \lambda^{2}} \sin 2 \phi + \alpha' \partial_{t} \phi \nonumber \\
    &+ \frac{U}{2\lambda^2} \sech^2\frac{x}{\lambda}(\phi - \phi_c)= 0,
\end{align}
where $\alpha'=\alpha v_m/a$. To solve Eq.~\eqref{eq:dynamical_eq_re}, we need two initial conditions, $\phi(x,0)$ and $\dot{\phi}(x,0)$. Since the laser pulse duration is much shorter than the characteristic time of the Néel vector dynamics, it is natural to set $\phi(x,0)$ as $\phi_0(x)$ obtained in Eq.~\eqref{eq:static} while $\dot{\phi}(x,0)$ (the initial angular velocity of $\mathbf{n}$) essentially follows the spatial profile of the laser pump. For a normally incident laser carrying a perpendicular spin polarization, $\dot{\phi}(x,0)$ is proportional to the amount of injected spins by the pump and its sign is determined by the pump helicity because at the $t=0$ instant (right after the pump is off), $\mathbf{m}\sim\mathbf{n}\times\dot{\mathbf{n}}$~\cite{chengUltrafastSwitching2015}. With these considerations, we can set
\begin{subequations}
\begin{align}
 \phi(x, 0)&=\phi_0(x), \\
 \dot{\phi}(x, 0)&= \pm\omega_0\exp[-2(x - x_{0})^{2}/r_{0}^{2}],
\end{align}
\end{subequations}
where $\pm$ corresponds to opposite helicities, $x_0$ marks the distance of the pump center from the origin, and $r_0$ is the radius of the Gaussian beam.

We numerically solved Eq.~\ref{eq:dynamical_eq_re} using the forward Euler method. The parameters used in generating Fig.~\ref{fig:m4}c and~d are: $v_{m} = \SI{0.2}{\um\per\ps}$, $\lambda = \SI{2.86}{\um}$, $\alpha' = \SI{3.58e-2}{\per\ps}$, $U = \SI{1.5e-3}{\um^2\per\ps\squared}$, and $\omega_0 = \SI{0.3}{\degree\per\ps}$.

\section{Acknowledgments}
\begin{acknowledgments}
We acknowledge helpful conversations with Jonathan Curtis and Eugene Demler. D.H. acknowledges support for time-resolved SHG measurements from the Brown Investigator Award, a program of the Brown Science Foundation, as well as the Institute for Quantum Information and Matter (IQIM), an NSF Physics Frontiers Center (PHY-2317110). D.H. acknowledges support for instrumentation from the David and Lucile Packard Foundation. K.L.S. acknowledges a Caltech Prize Postdoctoral Fellowship. R.C. and H.Z. were supported by the Air Force Office of Scientific Research under Grant No. FA9550-19-1-0307. The work at Stanford and SLAC (crystal growth and sample characterization) was supported by the U.S. Department of Energy (DOE), Office of Science, Basic Energy Sciences, Materials Sciences and Engineering Division, under contract DE-AC02-76SF00515.
\end{acknowledgments}

\section{Author Contributions}
K.L.S. and D.H. conceived the experiment. K.L.S. and D.V.B. performed the experiments. K.L.S. analyzed data. H.Z and R.C. led theoretical modeling and interpretation. C.R.R. and Y.S.L. synthesized and characterized samples. K.L.S. and D.H. wrote the paper with input from all authors.

\section{Competing Interests}

The authors declare no competing interests.

\newpage
\onecolumngrid

\begin{center}
    \textbf{\large Supplementary Information for \\``High-speed antiferromagnetic domain walls driven by coherent spin waves"}
\end{center}

\setcounter{equation}{0}
\setcounter{figure}{0}
\setcounter{table}{0}
\setcounter{page}{1}
\makeatletter
\renewcommand{\theequation}{S\arabic{equation}}
\renewcommand{\thefigure}{S\arabic{figure}}
\renewcommand{\bibnumfmt}[1]{[S#1]}
\renewcommand{\citenumfont}[1]{S#1} 

\section{S1. Rotational anisotropy of SHG on a DW}
\label{sec:s1}
\begin{figure}[!htb]
\includegraphics{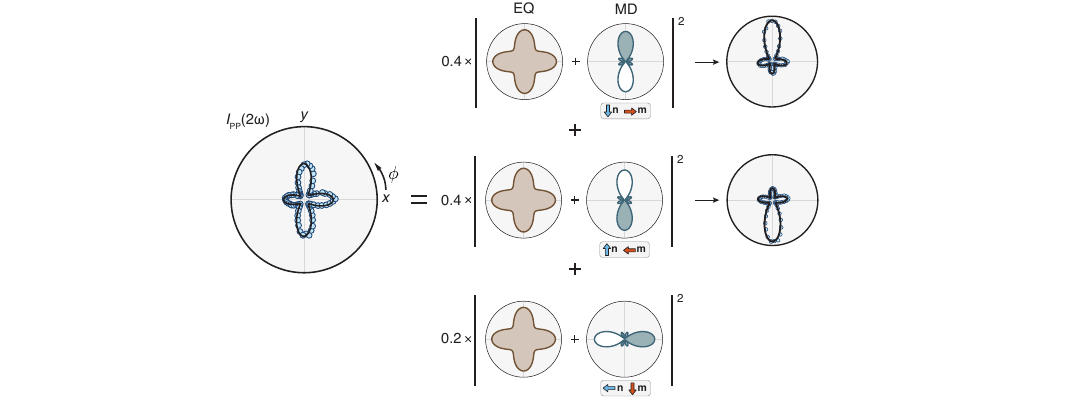}
\caption{\label{fig:s1} SHG rotational anisotropy polar plot (with \textit{P}-polarized input and output electric field polarization). The solid line is a fit to a three-domain averaged intensity, where each domain has SHG arising from a coherent superposition of electric quadrupole and magnetic dipole SHG processes, which are illustrated on the right. Filled and white lobes indicate opposite phase. The RA patterns on the right show the single domain data and fits.}
\end{figure}

\noindent Previous work showed that the SHG rotational anisotropy of \ce{Sr2Cu3O4Cl2} below \tnone{} can be described by a coherent superposition of electric quadrupole and magnetic dipole processes, which are sensitive to the crystallographic and magnetic order respectively [50]. We model the SHG rotational anisotropy pattern on a domain wall (DW) by the three-domain averaged intensity $I(2\omega,\phi)=\sum_{d=1,2,3} f_d|A\hat{e}_i^{\textrm{out}}\chi_{ijkl}^{\textrm{EQ}(i)}(\phi)\hat{e}_j^{\textrm{in}}q_k\hat{e}_l^{\textrm{in}} +A\hat{e}_i^{\textrm{out}}\chi_{ijk,d}^{\textrm{MD}(c)}(\phi)\hat{e}_j^{\textrm{in}}\epsilon_{klm}q_l\hat{e}^{\textrm{in}}_m|^2 I(\omega)^2$, where $\chi_{ijkl}^{\textrm{EQ}(i)}(\phi)$ is the $i$-type electric quadrupole susceptibility tensor transformed into the frame of the rotated scattering plane, $\chi_{ijk,d}^{\textrm{MD}(c)}(\phi)$ is the domain-dependent $c$-type magnetic dipole susceptibility tensor transformed into the frame of the rotated scattering plane, $\vec{q}$ the wavevector of incident light, $\hat{e}$ is the polarization of incoming fundamental or outgoing second-harmonic light, $I(\omega)$ is the intensity of the fundamental beam, $A$ is a constant that depends on the polarization geometry, $f_d$ accounts for the different domain fractions under the beam spot, and $\epsilon_{klm}$ is the Levi-Civita symbol. We first fit an SHG pattern from a single domain region (Fig.~\ref{fig:s1}, top right, using $f_2=f_3=0$) to extract the components for $\chi_{ijkl}^{\textrm{EQ}(i)}$ and $\chi_{ijk,1}^{\textrm{MD}(c)}$, which also gives $\chi_{ijk,2}^{\textrm{MD}(c)}$ and $\chi_{ijk,3}^{\textrm{MD}(c)}$ after \ang{180} and \ang{270} rotations respectively. We then fit the DW SHG pattern by only varying the relative domain intensities $f_1$, $f_2$, and $f_3$. As shown in Fig.~\ref{fig:s1}, this process yields an excellent fit to our observed SHG pattern and demonstrates that the DW is Néel-type with $\mathbf{n}$ oriented along $-x$ at the DW center. When the laser is centered on the DW, we find that $\SI{\sim 20}{\percent}$ of the SHG intensity arises from the DW itself (and $\SI{\sim 40}{\percent}$ from each of the two antiphase domains). Because the SHG intensity scales with the square of the magnetization, this implies that the DW fills $\SI{\sim 26}{\percent}$ of the \SI{40}{\um} diameter laser spot. The DW width can then be estimated as $0.26\pi (\SI{20}{\um})^2/(\SI{40}{\um})\approx \SI{8.2}{\um}$, which is in excellent agreement with the \SI{\sim 9}{\um} width extracted by directly fitting the DW profile.

\newpage
\section{S2. Pump laser position dependence}
\label{sec:s2}

\begin{figure}[!htb]
\includegraphics{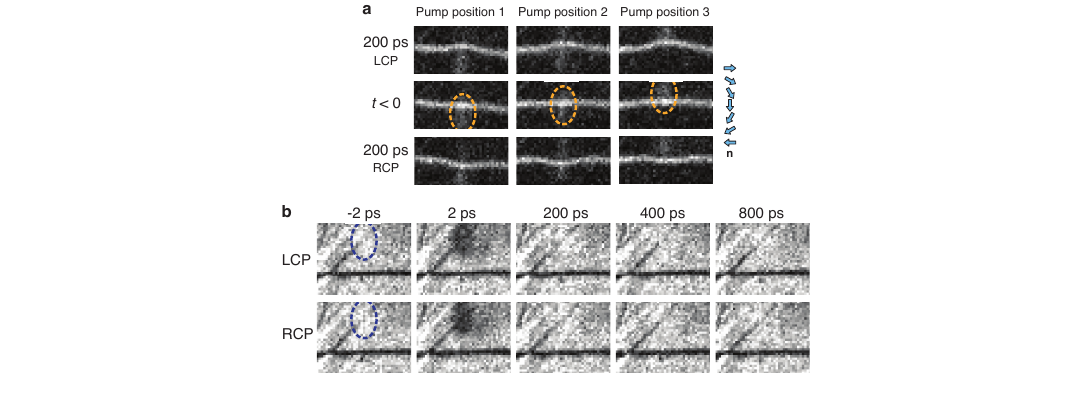}
\caption{\label{fig:s2} \textbf{a}, SHG images of a DW for different pump laser positions (FWHM indicated by the dashed ovals). Images are displayed at $t < 0$ (center row) and $t=\SI{200}{\ps}$ for left (top row) and right (bottom row) pump circular polarization. The DW configuration (Néel vector orientations across the DW) is shown on the right. The vertical scale of each image is \SI{20}{\um}. \textbf{b}, SHG images of an antiphase DW at different pump-probe time delays for left and right circular polarization when the pump laser is located completely off the DW. The SHG excitation polarization is orthogonal to the polarization used in \textbf{a}, so dark and bright regions are swapped. This allows one to clearly see the pumped region through the ultrafast melting of the magnetic order at $t=\SI{2}{\ps}$. The vertical scale of each image is \SI{30}{\um}.}
\end{figure}

\noindent Figure~\ref{fig:s2} shows how the light-induced DW motion depends on the position of the pump laser spot relative to the DW. For a fixed pump helicity, the DW always moves in the same direction, regardless of whether the pump laser spot is centered on the wall or spatially offset above or below it (Fig.~\ref{fig:s2}a), which rules out simple laser heating effects as the underlying mechanism of DW motion. When the pump is offset from the DW, the DW motion is slightly more pronounced in the direction of higher pump intensity. This is not surprising since greater motion is expected with higher spin wave amplitude (Fig.~\ref{fig:s5}). The asymmetry is also seen in the simulations (Fig.~\ref{fig:s8}). However, if the pump beam does not directly excite the DW, as in Fig.~\ref{fig:s2}b, we do not observe any motion. This suggests that the spin waves do not propagate appreciably outside of the pumped region.

\newpage
\section{S3. Pump scattering plane angle dependence}
\label{sec:s3}
\begin{figure}[!htb]
\includegraphics{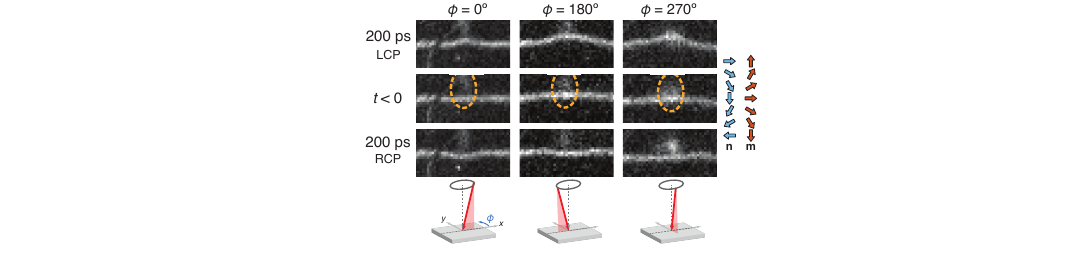}
\caption{\label{fig:s3} \textbf{a}, SHG images of a DW for different pump laser scattering plane angles, $\phi=\ang{0}$, $\phi=\ang{180}$, and $\phi=\ang{270}$ (schematically shown below each data column). Images are displayed at $t < 0$ (center row) and $t=\SI{200}{\ps}$ for left (top row) and right (bottom row) circular polarization. The angle of incidence is \ang{\sim 10}. The DW configuration, including the weak magnetic moments, is shown on the right. The vertical scale of each image is \SI{20}{\um}.}
\end{figure}

\noindent The inverse Faraday effect generates an effective magnetic field along the wavevector of circularly polarized pump light. For an oblique incidence pump pulse, this can create a transient in-plane magnetic field that couples to the weak in-plane magnetization of \ce{Sr2Cu3O4Cl2} and thereby induce DW motion. In this case, one would expect the DW motion to reverse with the opposite pump scattering plane angle. However, as shown in Fig.~\ref{fig:s3}, for a given pump helicity, the DW always moves in the same direction, independent of the scattering plane angle. Therefore, the light-induced motion cannot arise from a transient in-plane opto-magnetic field.

\newpage
\section{S4. Motion under time-reversed configuration}
\label{sec:s4}
\begin{figure}[!htb]
\includegraphics{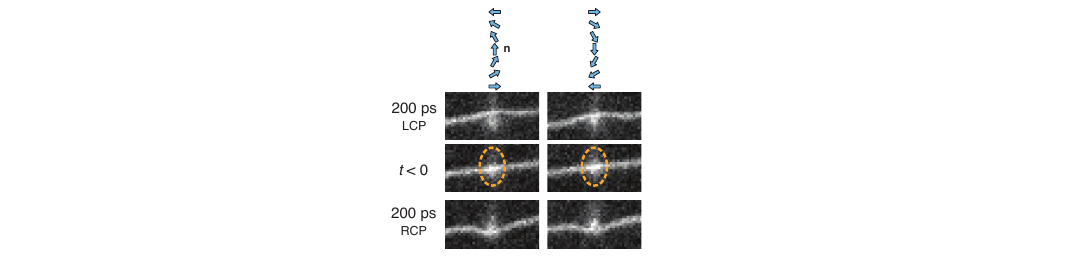}
\caption{\label{fig:s4} SHG images of a DW for two different DW configurations, which are the time reversal of one another. Images are displayed at $t < 0$ (center row) and $t=\SI{200}{\ps}$ for left (top row) and right (bottom row) circular polarization. The vertical scale of each image is \SI{20}{\um}.}
\end{figure}

\noindent Figure~\ref{fig:s4} shows that time-reversing the DW configuration (i.e., flipping all spins by \ang{180}) does not affect the light-induced motion. The direction of the Néel vector and weak magnetic moment within the DW does not matter. Instead, the direction of light-induced wall motion is fully determined by the pump photon helicity and the sign of the DW winding number (whether the spins rotate clockwise or anti-clockwise across the wall).

\newpage
\section{S5. Pump fluence dependence}
\label{sec:s5}

\begin{figure}[!htb]
\includegraphics{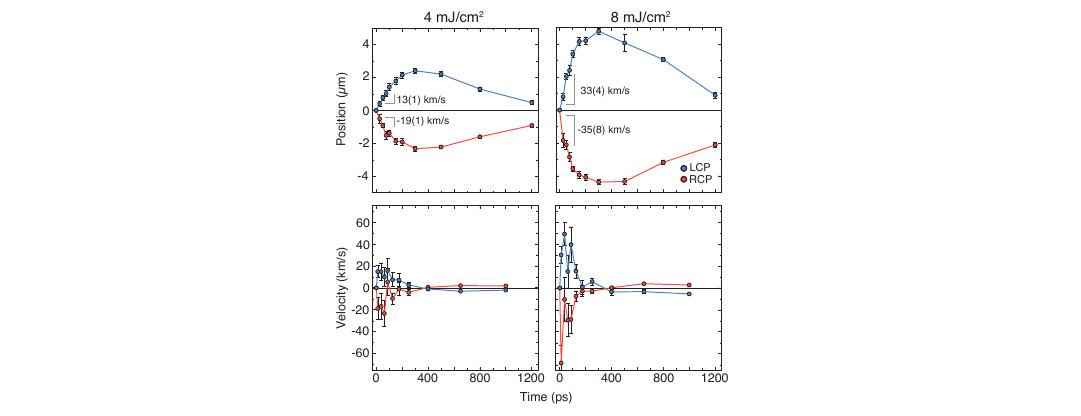}
\caption{\label{fig:s5} Fitted position and velocity of the DW center (at the location of maximum movement) over time for different pump helicities at \SI{4}{\milli\joule\per\cm^2} (left two plots) and \SI{8}{\milli\joule\per\cm^2} (right two plots). The extracted velocity from a linear fit to the first \SI{100}{\ps} of data in the position-versus-time plots is also displayed. The DW was located in a different sample position compared to the data in Fig.~4. The pump laser was carefully centered on top of the DW so that changing the fluence did not shift the equilibrium DW position. The fit uncertainties are given as 1 standard deviation.}
\end{figure}

\noindent Figure~\ref{fig:s5} shows the extracted DW dynamics for two different pump fluences. In both the low and high fluence cases, the DW accelerates outward in the first \SI{\sim 100}{\ps}, then slows down and eventually reverses direction. The timescale of this outward acceleration, and the fact that it does not change with fluence, is consistent with our picture of light-induced spin waves driving the motion. The initial in-plane Néel vector rotation that occurs in the light-induced spin wave also occurs over \SI{\sim 100}{\ps}, as seen in Fig.~3e, and this spin wave frequency is not affected by the pump fluence. On the other hand, the spin wave amplitude increases with pump fluence, which explains why we observe an increase in the maximum DW position and maximum velocity when the fluence is increased. Further experimental studies are required to understand exactly how the DW velocity and maximum position scale with fluence. We note that the relationship may be complicated by the fact that higher pump laser fluence induces additional local heating that changes the DW energy landscape, which can affect the dynamics.

\newpage
\section{S6. Additional DW dynamics simulations}
\label{sec:s7}

\begin{figure}[!htb]
\includegraphics{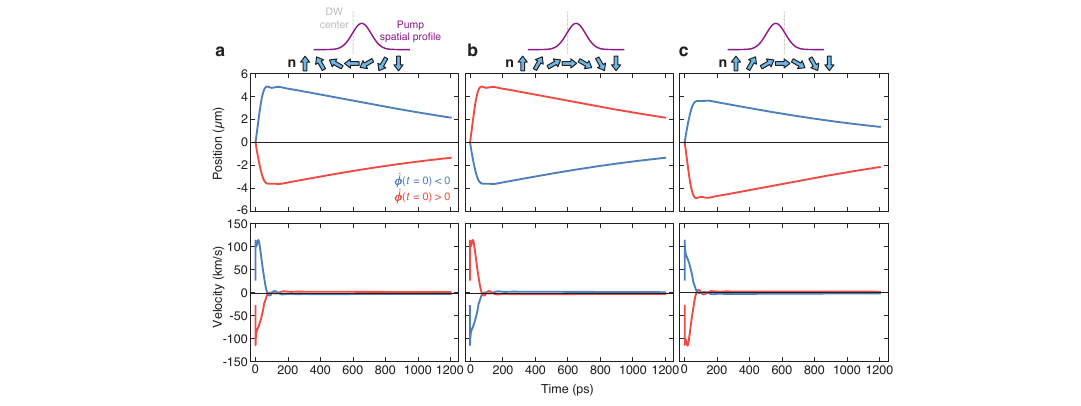}
\caption{\label{fig:s7} \textbf{a}, Simulated time dependence of the DW position and velocity for $w=-1/2$ and pump position spatially offset to the right of the DW. \textbf{b}, Simulated time dependence of the DW position and velocity for $w=1/2$ and pump position spatially offset to the right of the DW. \textbf{c}, Simulated time dependence of the DW position and velocity for $w=1/2$ and pump position spatially offset to the left of the DW.}
\end{figure}

\noindent Figure~\ref{fig:s7} shows DW dynamics simulations for different winding numbers and pump locations. In particular, comparing Fig.~\ref{fig:s7}a (where $w=-1/2$) and Fig.~\ref{fig:s7}b (where $w=1/2$), we see that the dynamics are reversed for a chosen sign of $\dot{\phi}(t=0)$. Furthermore, the effect of the pump laser position relative to the DW center is seen by comparing Fig.~\ref{fig:s7}b, where the pump profile is positively offset from the DW, to Fig.~\ref{fig:s7}c, where the pump profile is negatively offset from the DW. The direction of DW motion does not depend on the pump laser position, but the slight asymmetry between the $\dot{\phi}(t=0)>0$ and $\dot{\phi}(t=0)<0$ cases does. These findings are consistent with the experimental observations.

\newpage
\section{S7. DW motion under spatially uniform pump}
\label{sec:s8}

\begin{figure}[!htb]
\includegraphics{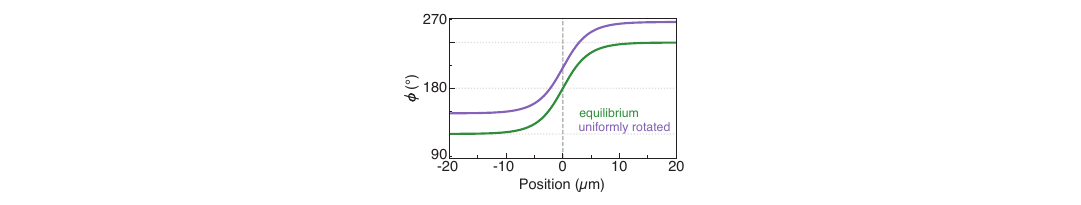}
\caption{\label{fig:s8} Antiphase DW profile at equilibrium and after a uniform clockwise rotation of every spin.}
\end{figure}

\noindent A spatially uniform rotation of spins merely shifts the global phase of a DW rather than shifting its position, as DW motion requires non-uniform spin rotation to accommodate the DW's inherent spatial inhomogeneity. This is seen in Fig.~\ref{fig:s8}a, where we compare a DW at equilibrium and immediately after excitation by a uniform stimulus. Although the spins are out of equilibrium with a global phase shift, the DW center remains fixed at $x=0$. This observation raises a fundamental question: how does DW motion emerge when subjected to a uniform stimulus, or more generally, when the pump's spatial profile is mismatched to the DW profile? To provide physical insight on this question, here we consider DW motion under the influence of spatially uniform stimulus as a simple case of spatial profile mismatch.

To simplify the theoretical analysis without losing the essential physics, we drop the dissipation term, trapping potential, and coefficients before each term from the dynamical equation (Eq.~(6) in the Methods section) and keep a minimal model for the DW:
\begin{equation}
    \partial_t^2 \phi - \partial_x^2 \phi - \sin 2\phi = 0.
    \label{eq:1}
\end{equation}
The static DW profile without stimulus, $\phi_0(x)$, is obtained by solving Eq.~\eqref{eq:1} with $\partial_t \phi = 0$, so it satisfies
\begin{equation}
    \partial_x^2 \phi_0 + \sin 2\phi_0 = 0.
    \label{eq:2}
\end{equation}
After the DW is exposed to stimulus, the DW profile can be decomposed as
\begin{equation}
    \phi(x,t) = \phi_0(x) + \Delta \phi(x,t),
    \label{eq:3}
\end{equation}
where $\Delta\phi(x,t)$ is a small stimulus-induced dynamical part on top of the static background. For a uniform stimulus, $\Delta \phi(x,t=0) = \Delta \phi(t=0) \equiv \Delta \phi(0)$ and $\partial_t \Delta \phi(x,t=0) = \partial_t \Delta \phi(t=0) \equiv \partial_t \Delta \phi(0)$, where $\Delta \phi(0)$ and $\partial_t \Delta \phi(0)$ serve as our initial conditions for Eq.~\eqref{eq:1}. Substituting Eq.~\eqref{eq:3} into the dynamical equation (Eq.~\eqref{eq:1}) and noting that $\phi_0$ is time-independent ($\partial_t^2 \phi_0 = 0$), we get
\begin{equation}
    \partial_t^2 \Delta \phi - \partial_x^2 \phi_0 - \partial_x^2 \Delta \phi - \sin 2(\phi_0 + \Delta \phi) = 0.
    \label{eq:4}
\end{equation}
For small $\Delta \phi$, we can perform a Taylor expansion of the sine term as $\sin 2(\phi_0 + \Delta \phi) \approx \sin 2\phi_0 + 2(\cos 2\phi_0)\Delta \phi$. Using Eq.~\eqref{eq:2}, we therefore simplify Eq.~\eqref{eq:4} to:
\begin{equation}
    \partial_t^2 \Delta \phi - \partial_x^2 \Delta \phi - 2(\cos 2\phi_0)\Delta\phi = 0.
    \label{eq:5}
\end{equation}
At $t=0$, since the stimulus is uniform, $\partial_x^2 \Delta \phi(t=0) = 0$, then Eq.~\eqref{eq:5} becomes
\begin{equation}
    \partial_t^2 \Delta \phi(t=0) = 2(\cos 2\phi_0)\Delta \phi(0).
    \label{eq:6}
\end{equation}
The left-hand side represents the ``angular acceleration" of spins at $t=0$ under the stimulus. The right-hand side is spatially non-uniform because the static DW profile $\phi_0(x)$ varies with position despite the uniform stimulus $\Delta \phi(0)$. Consequently, this non-uniform angular acceleration leads to a non-uniform rotation of the spins in the DW, which is ultimately necessary for the DW motion. This analysis does not contradict the previous statement that a spatially uniform rotation of spins cannot move a DW. Immediately after the initial stimulus, the DW center remains at $x=0$, and it is the follow-up evolution that creates the required inhomogeneity.

In the case where $\Delta \phi(0) = 0$ but $\partial_t \Delta \phi(0) \neq 0$, we have $\partial_t^2 \Delta \phi(t=0) = 0$ according to Eq.~\eqref{eq:6}, meaning spins initially rotate uniformly without acceleration. However, after a small time $\Delta t$, $\Delta \phi(x, t=\Delta t) = \Delta t \times \partial_t \Delta \phi(0) \neq 0$ is still uniform, and we can repeat our analysis with the same result. This treatment is only valid at the beginning of the evolution and assumes a small stimulus relative to the DW profile.

\begin{figure}[h]
\includegraphics{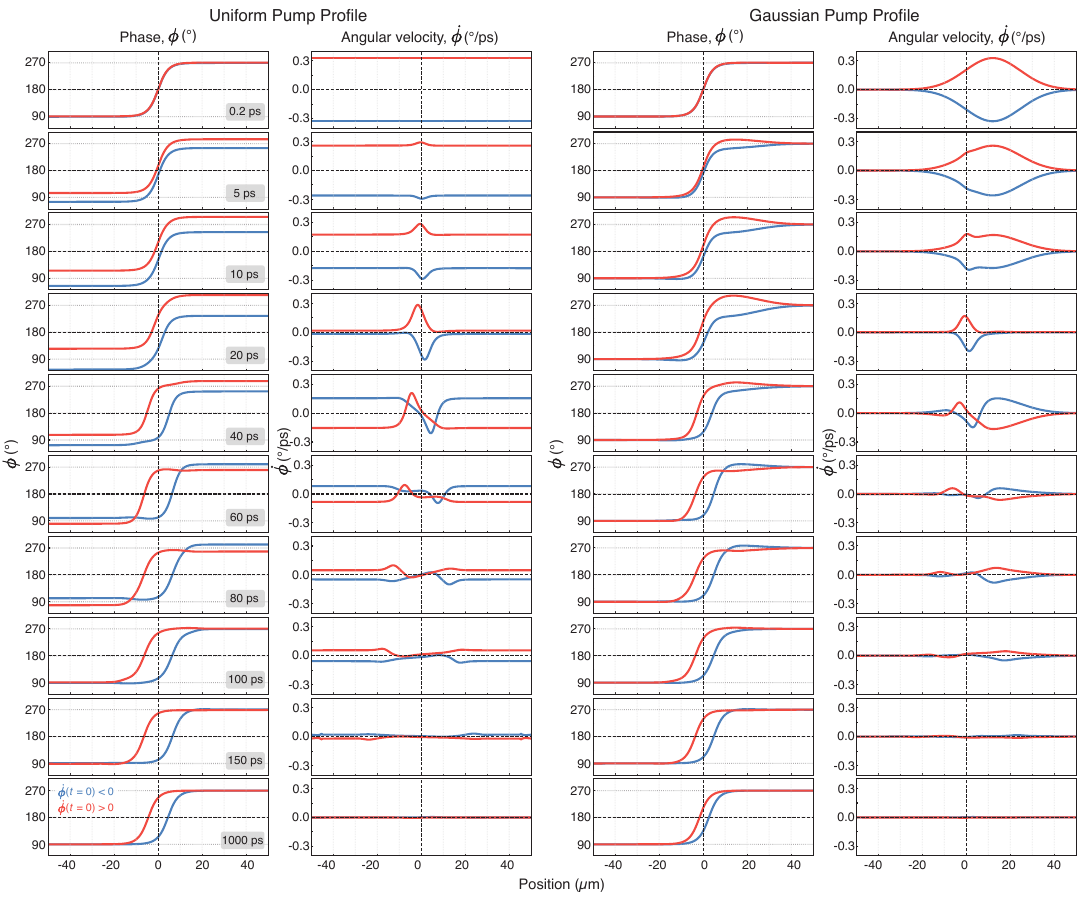}
\caption{\label{fig:s9} Simulated time evolution of the DW phase and angular velocity profiles after excitation by a spatially uniform (left) or Gaussian (right) pump profile. The Gaussian pump case uses the same width and spatial offset as the other simulations in the text. For the uniform pump case, a top-hat pulse shape with \SI{75}{\um} width was used. The simulation ranged from \SIrange{-100}{100}{\um}. Each row corresponds to the same time indicated in the leftmost column.}
\end{figure}

To validate our understanding, we plot the simulated phase and angular velocity profiles after uniform and Gaussian pump excitation for the full dynamical equation in the main text (Fig.~\ref{fig:s9}). While the initial angular velocity follows the profile of the pump pulse (either uniform or Gaussian), it develops additional inhomogeneity soon afterward, arising from the non-uniform angular acceleration of the spins due to the DW profile.

\end{document}